# Intra-unit-cell electronic nematicity of the high-$T_c$ copper-oxide pseudogap states

*M. J. Lawler[1,2], *K. Fujita[2,3,4], Jhinhwan Lee[2,3,5], A.R. Schmidt[2,3], Y. Kohsaka[6], Chung Koo Kim[2,3], H. Eisaki[7], S. Uchida[4], J.C. Davis[2,3,8], J.P. Sethna[2], and Eun-Ah Kim[2]

[1]Department of Physics, Applied Physics and Astronomy, Binghamton University, Binghamton, NY 13902-6000, USA. [2]Laboratory for Atomic and Solid State Physics, Department of Physics, Cornell University, Ithaca, NY 14853, USA. [3]Condensed Matter Physics and Materials Science Department, Brookhaven National Laboratory, Upton, NY 11973, USA. [4]Department of Physics, University of Tokyo, Bunkyo-ku, Tokyo 113-0033, Japan. [5]Dept. of Physics, Korea Advanced Institute of Science and Technology, 373-1 Guseong-dong, Yuseong-gu, Daejeon 305-701, Korea. [6]Magnetic Materials Laboratory, RIKEN, Wako, Saitama 351-0198, Japan. [7]Institute of Advanced Industrial Science and Technology, Tsukuba, Ibaraki 305-8568, Japan. [8]School of Physics and Astronomy, University of St. Andrews, North Haugh, St. Andrews, Fife KY16 9SS, Scotland.
* These authors contributed equally to this work

**In the high-transition-temperature (high-$T_c$) superconductors the pseudogap phase becomes predominant when the density of doped holes is reduced[1]. Within this phase it has been unclear which electronic symmetries (if any) are broken, what the identity of any associated order parameter might be, and which microscopic electronic degrees of freedom are active. Here we report the determination of a quantitative order parameter representing intra-unit-cell nematicity: the breaking of rotational symmetry by the electronic structure within $CuO_2$ unit cell. We analyze spectroscopic-imaging scanning tunneling microscope images of the intra-unit-cell states in underdoped $Bi_2Sr_2CaCu_2O_{8+\delta}$ and, using two independent evaluation techniques, find evidence for electronic nematicity of the states close to the pseudogap energy. Moreover, we demonstrate directly that these phenomena arise from electronic differences at the two oxygen sites within each unit cell. If the characteristics of the pseudogap seen here and by other techniques all have the same** **microscopic origin, this phase involves weak magnetic states at the O sites that break 90º –rotational symmetry within every $CuO_2$ unit cell.**

The pseudogap phase emerges in the $CuO_2$ plane (Fig. 1a) upon removal of electrons from the O atoms of the parent 'charge-transfer'[2] Mott insulator (Fig. 1b). At low hole-density $p$, two fundamentally distinct types of electronic excitations are observed[3] (Fig. 1c). Here the lower-energy branch of excitations $\Delta_0$ appears to be associated with the maximum energy of Cooper pairing[4]. The higher-energy branch labeled $\Delta_1$ is the spectroscopic 'pseudogap' because the same energy gap exists in both the superconducting and pseudogap phases (Fig. 1b). Moreover, $\Delta_1$ tracks $T^*$, the temperature at which pseudogap phenomenology appears, as a function of doping, implying that $\Delta_1$ is the energy gap associated with any electronic symmetry breaking at $T^*$. Therefore the question of which symmetries are broken at $T^*$ is equivalent to that of which broken symmetries the excitations with energy $\Delta_1$ exhibit.

The 'pseudogap states' at energy $\omega \approx \pm \Delta_1$ can be examined directly using tunnelling spectroscopy. Figure 1d shows the evolution of spatially averaged conductance spectra[4] with the doping dependence of $\pm \Delta_1$ indicated by blue dashed curves. Figure 2a and b shows atomically resolved electronic structure images with $\omega \approx \Delta_1$ (refs 5 and 6) acquired at $T < T_c$ (Fig. 2a) and $T > T_c$ (Fig. 2b) for an underdoped $Bi_2Sr_2CaCu_2O_{8+\delta}$ sample with $T_c \approx 35$ K. The insets show that their Fourier transforms are dominated by four inequivalent non-dispersive wavevectors labelled $\mathbf{Q}_x$, $\mathbf{Q}_y$, $\mathbf{S}_x$ and $\mathbf{S}_y$. These phenomena are indistinguishable above and below $T_c$ (refs 5 and 6 and Fig. 2), so the spatial symmetries of the pseudogap states can be established to the highest precision by using the $T \approx 0$ experiments. A striking feature of such pseudogap state images in $Bi_2Sr_2CaCu_2O_{8+\delta}$ (refs 5–8) and $Ca_{2-x}Na_xCuO_2Cl_2$ (ref. 5) is the apparent local breaking of 90º rotation ($C_4$) symmetry at the unit-cell scale. Lattice-translation symmetry is also broken locally at the nanoscale, yielding broad peaks around $\mathbf{S}_x$ and $\mathbf{S}_y$ (refs 5 and 6).

Theoretical analyses of such images have posited disordered smectic electronic liquid crystals or stripes[9,10].

Indeed, from the perspective of symmetry, the spatial arrangements of the pseudogap states[5–8] exhibit similarities to those of liquid crystals. Electronic liquid crystals have been defined[11–17] to arise when the electronic structure breaks spatial symmetries of the crystal lattice (any resultant lattice symmetry changes may or may not be detectable[16]). Nematic phases preserve the (lattice) translational symmetry and break the rotational (point group) symmetry, whereas smectic and striped phases break both[10–17]. Correlated electronic systems exhibiting evidence for electronic nematicity include $Sr_3Ru_2O_7$ (ref. 18) and $CaFe_{1.94}Co_{0.06}As_2$ (ref. 19). Microscopic models for both systems posit nematicity due to a $C_4$ symmetry breaking which renders the $d_{xz}$ and $d_{yz}$ orbitals of each Ru or Fe atom inequivalent [20–23]. For copper oxides, the local 180° rotational ($C_2$) symmetry observed in the pseudogap states[5,6,8] implied the possibility of electronic nematicity but with contributions from multiple sites within the $CuO_2$ cell (shown schematically in Fig. 2c).

Here we focus on $C_4$ to $C_2$ symmetry reduction in sub-unit-cell resolution images of the copper-oxide pseudogap states[5,6,8]. We introduce a general measure of electronic nematicity in any image $M(\mathbf{r})$:

$$O_n[M] = \frac{1}{2}[\tilde{M}(\mathbf{Q}_y) - \tilde{M}(\mathbf{Q}_x) + \tilde{M}(-\mathbf{Q}_y) - \tilde{M}(-\mathbf{Q}_x)] = \mathrm{Re}\,\tilde{M}(\mathbf{Q}_y) - \mathrm{Re}\,\tilde{M}(\mathbf{Q}_x). \quad (1)$$

Here $\tilde{M}(\mathbf{q}) = \mathrm{Re}\,\tilde{M}(\mathbf{q}) + i\,\mathrm{Im}\,\tilde{M}(\mathbf{q})$, the complex-valued two-dimensional Fourier transform of any $M(\mathbf{r})$ with its origin at a Cu site, is evaluated at inequivalent Bragg peaks $\mathbf{Q}_x = (2\pi/a_0)\,\mathbf{x}$ and $\mathbf{Q}_y = (2\pi/a_0)\,\mathbf{y}$, and $a_0$ is the unit-cell dimension. Figure 1a illustrates why only $\mathrm{Re}\,\tilde{M}(\mathbf{q})$ enters the definition of $O_n[M]$ by showing it in phase with the Cu lattice (even about each Cu site). By focusing on the Bragg peaks, $O_n[M]$ locks onto only those electronic phenomena with the same spatial periodicity as the crystal and then quantifies the $C_4$ symmetry breaking. Importantly, this generic order parameter in equation (1) will be widely applicable to data from other techniques. Successful application of equation (1) for spectroscopic-imaging scanning tunneling microscopy (STS) images requires (1) a highly accurate registry of each Cu site location in $M(\mathbf{r})$ to satisfy the extreme sensitivity of $O_n[M]$ to the phase of $\tilde{M}(\mathbf{Q}_{x,y})$, and (2) that $M(\mathbf{r})$ has the sub-unit-cell resolution without which $O_n[M]$ will be identically zero. $O_n[M]$ would also be identically zero if only the Cu sites $\mathbf{R} = (ma_0, na_0)$ (with $m$, $n$ integers) contribute to $M(\mathbf{r})$ because then $\tilde{M}(\mathbf{Q}_x) \equiv \tilde{M}(\mathbf{Q}_y)$. In contrast, if all three atomic sites within each $CuO_2$ unit cell contribute to $M(\mathbf{r})$, then:

$$\tilde{M}(\mathbf{Q}_x) = \overline{M}_{Cu} - \overline{M}_{O_x} + \overline{M}_{O_y}$$
$$\tilde{M}(\mathbf{Q}_y) = \overline{M}_{Cu} + \overline{M}_{O_x} - \overline{M}_{O_y} \qquad (2)$$
$$O_n[M] = 2(\overline{M}_{O_x} - \overline{M}_{O_y})$$

where $\overline{M}_{Cu}$, $\overline{M}_{O_x}$ and $\overline{M}_{O_y}$ are the average of $M(\mathbf{r})$ at Cu, $O_x$ and $O_y$ sites respectively. The last line follows from equation (1). Hence detection of $O_n[M] \neq 0$ would imply that an inequivalence exists between the electronic structure at the $O_x$ and $O_y$ sites, as shown schematically in Fig. 2c.

Standard differential conductance mapping of $dI/dV(\mathbf{r}, \omega = eV) \equiv g(\mathbf{r}, \omega)$ to image the local density of states $N(\mathbf{r}, \omega)$ is challenging in strongly underdoped copper-oxides[5–8] because of the intense electronic heterogeneity. However, imaging the ratio $Z(\mathbf{r}, \omega) \equiv \frac{g(\mathbf{r}, +\omega)}{g(\mathbf{r}, -\omega)} \equiv \frac{N(\mathbf{r}, +\omega)}{N(\mathbf{r}, -\omega)}$ avoids these severe systematic errors[5–8] (Supplementary Information section I). Another challenge is the random nanoscale variations of $\Delta_1(\mathbf{r})$; this problem can be mitigated[8] by defining a reduced energy scale $e \equiv \omega / \Delta_1(\mathbf{r})$ which measures the energy at each position $\mathbf{r}$ with respect to the pseudogap magnitude $\Delta_1(\mathbf{r})$ at that position, thus ensuring that the pseudogap states occur at $e \cong 1$ in all $Z(\mathbf{r}, e)$ data[8]. Using these techniques, a clearer picture of the two basic classes[3] of electronic excitations in copper oxides has emerged[5–8]. First, for all energies below $\omega \approx \Delta_0$ (a slowly doping dependent energy[8] indicated by the red dashed line in Fig. 1d) the electronic excitations are demonstrably[5–8] Bogoliubov quasiparticles. Indeed, in both the superconducting[5–8] and underdoped pseudogap[6] phases, these dispersive low-energy states exhibit only the symmetries of a $d$-wave superconductor. The $\omega \approx \Delta_1$ pseudogap states are profoundly different, being non-dispersive[5–8] and locally breaking crystal point-group symmetries. The challenge is therefore to identify first which symmetries are broken by the $\omega \approx \Delta_1$ pseudogap states and then the microscopic degrees of freedom involved in any such symmetry breaking.

We explore the intra-unit-cell electronic nematicity of $Bi_2Sr_2CaCu_2O_{8+\delta}$ by applying equation (1) to $\tilde{Z}(\mathbf{Q}_x, e)$ and $\tilde{Z}(\mathbf{Q}_y, e)$, the complex-valued two-dimensional Fourier transforms of $Z(\mathbf{r}, e)$ evaluated at $\mathbf{Q}_x$ and $\mathbf{Q}_y$ (red circles in Fig. 2). The sub-unit-cell resolution $Z(\mathbf{r}, e)$ imaging was performed on multiple different underdoped $Bi_2Sr_2CaCu_2O_{8+\delta}$ samples with values of $T_c$ between 20 K and 55 K. The necessary registry of the Cu sites in each $Z(\mathbf{r}, e)$ is achieved by a picometre-scale realignment procedure that identifies a transformation to render the topographic image $T(\mathbf{r})$ perfectly $a_0$-periodic (Supplementary Information section II). We then apply the same transformation to the simultaneously acquired $Z(\mathbf{r}, e)$ to register all the electronic structure data to this ideal lattice.

Such a topographic image $T(\mathbf{r})$ is shown in Fig. 3a; the location of every Cu site is known with a precision of about 10 picometres. The inset compares the Bragg peaks of its real (in phase) Fourier components $\mathrm{Re}T(\mathbf{Q}_x)$ and $\mathrm{Re}T(\mathbf{Q}_y)$ and demonstrates that $\mathrm{Re}T(\mathbf{Q}_x) / \mathrm{Re}T(\mathbf{Q}_y) = 1$. Therefore $T(\mathbf{r})$ preserves the $C_4$ symmetry of the crystal lattice. In contrast, Fig. 3b shows that $Z(\mathbf{r}, e = 1)$ determined simultaneously with Fig. 3a seems to break various crystal symmetries locally[5–7]. The inset showing that $\mathrm{Re}\tilde{Z}(\mathbf{Q}_x, e) / \mathrm{Re}\tilde{Z}(\mathbf{Q}_y, e) \neq 1$ reveals a remarkable discovery: the pseudogap states break $C_4$ symmetry throughout Fig. 3b. Therefore, following equation (1), we define a normalized order parameter over the entire field of view (FOV) for intra-unit-cell electronic nematicity as a function of $e$:

$$O_n^Q(e) \equiv \frac{\operatorname{Re}\tilde{Z}(\mathbf{Q}_y,e) - \operatorname{Re}\tilde{Z}(\mathbf{Q}_x,e)}{\bar{Z}(e)} \qquad (3)$$

The plot of $O_n^Q(e)$ in Fig. 3c provides the second important discovery: the magnitude of $O_n^Q(e)$ is low for $e \ll 1$, only begins to grow near $e \approx \Delta_0 / \Delta_1$, and becomes well defined near $e \approx 1$ ($\omega \approx \Delta_1 \cong 100$ meV in this sample). Hence, the electronic nematicity is associated with the pseudogap states. The low value of $|O_n^Q(e)|$ for small $e$ is to be expected, because here the Bogoliubov quasiparticles are observed to respect the *d*-wave superconductor symmetries[4,5,8]. The rapid increase in $|O_n^Q(e)|$ is primarily of electronic origin, because the normalization factor (the spatial average of $Z(\mathbf{r},e)$) $\bar{Z}(e) \cong 1$ for all $e$. Finally, equivalent $O_n^Q(e)$ analyses of all samples studied, measured using two different STMs, show results in agreement with Fig. 3 (Supplementary Information section III), with the exception that the opposite sign of $O_n^Q(e)$ can also be observed.

Equation (2) predicts that the predominant contributions to nematicity must come from the O sites (Fig. 2c). To test this idea directly, we examine $Z(\mathbf{r},e)$ with sub-unit-cell resolution. Figure 3d shows the topographic image of a representative region from Fig. 3a; the locations of each Cu site $\mathbf{R}$ and of the two O sites within its unit cell are indicated. Figure 3e shows $Z(\mathbf{r},e)$ measured simultaneously with Fig. 3d with the same Cu and O site labels. To quantify the $\mathbf{r}$-space electronic nematicity we define the $\mathbf{r}$-space equivalent of equation (3):

$$O_n^R(e) = \sum_{\mathbf{R}} \frac{Z_x(\mathbf{R},e) - Z_y(\mathbf{R},e)}{\bar{Z}(e)N} \qquad (4)$$

Here $Z_x(\mathbf{R},e)$ is the magnitude of $Z(\mathbf{r},e)$ at the O site $a_0/2$ along the *x*-axis from $\mathbf{R}$ while $Z_y(\mathbf{R},e)$ is the equivalent along the *y*-axis, and $N$ is the number of unit cells. Hence $O_n^R(e)$ counts only the O site contributions. Figure 3e shows the calculated value of $O_n^R(e)$ from the same FOV as Fig. 3a and b. It is obviously in good agreement with $O_n^Q(e)$, remaining at low amplitude for small $e$ and then rapidly becoming established near the pseudogap states at $e \approx 1$. Thus, the electronic nematicity discovered using $O_n^Q(e)$ indeed derives primarily from the inequivalence of electronic structure at the two O sites within each unit cell. The energy dependence of the average unit-cell electronic structure (Supplementary Information section IV) shows directly that although electronic changes are undetectable at the Cu sites, they are observable at the O sites; this is an explicit demonstration that electronic nematicity in $O_n^Q(e)$ is associated with the O sites and the pseudogap energy. All these observations have been confirmed, using both independent measures $O_n^Q(e)$ and $O_n^R(e)$, in the other samples (Supplementary Information section III). Moreover, other possible sources of systematic error including tip-shape, STM head geometry, scan direction, and sub-atomic register of $Z(\mathbf{r}, e)$ to $T(\mathbf{r})$ have been ruled out positively for every tip/sample combination reported herein (Supplementary Information section V).

Evidence that a predominant symmetry of the pseudogap states of $Bi_2Sr_2CaCu_2O_{8+\delta}$ is that of an intra-unit-cell nematic motivates examination of any smectic characteristics in $Z(\mathbf{r}, e)$, because a smectic can melt into a nematic[9–16]. Originally the searches for smectic patterns focused on $g(\mathbf{r}, \omega)$ modulations with $\mathbf{Q}_{1/4, x} \approx (1/4, 0) 2\pi / a_0$, $\mathbf{Q}_{1/4, y} \approx (0, 1/4) 2\pi / a_0$ (refs 9, 10 and 12). However, all $\mathbf{Q}_{1/4, x}$ and $\mathbf{Q}_{1/4, y}$ features at low energy are actually dispersive Bogoliubov quasiparticle interferences[6–8] and non-dispersive modulations with these $\mathbf{Q}$ values become very weak in $Z(\mathbf{r}, e \approx 1)$ (Figs 2 and 3). Instead, the strong nondispersive modulations at the pseudogap energy[5–8] exhibit wavevectors $\mathbf{S}_x \equiv (\sim 3/4, 0) 2\pi / a_0$, $\mathbf{S}_y \equiv (0, \sim 3/4) 2\pi / a_0$ (blue circles in Fig. 2 insets) the magnitudes $|\mathbf{S}_x|$ and $|\mathbf{S}_y|$ of which are weakly doping-dependent[8]. There are two important points here: (1) our definition of intra-unitcell nematicity using the Bragg peaks in equation (1) is independent of the $\mathbf{S}_x$, $\mathbf{S}_y$ (or any other) smectic modulations and, (2) although $\mathbf{S}_x$ and $\mathbf{S}_y$ might seem related to $\mathbf{Q}_{1/4, x}$ and $\mathbf{Q}_{1/4, y}$ by reciprocal lattice vectors, the two sets of wavevectors are actually inequivalent in a $Z(\mathbf{q}, e)$ derived from a $Z(\mathbf{r}, e)$ resolving the three atomic sites within the unit cell.

To examine smectic spatial organization, we next define a measure analogous to equation (3) of $C_4$ symmetry breaking in modulations with $\mathbf{S}_x$, $\mathbf{S}_y$ as:

$$O_s^Q(e) = \frac{\mathrm{Re}\,\tilde{Z}(S_y, e) - \mathrm{Re}\,\tilde{Z}(S_x, e)}{\bar{Z}(e)} \tag{5}$$

For all samples studied, $|O_s^Q(e)|$ is found to be very low and independent of energy (Fig. 4b). Obviously, $|O_s^Q(e)|$ is low at low $e$ because these states are dispersive Bogoliubov quasiparticles[6–8] but, more importantly, $|O_s^Q(e)|$ shows no tendency to become well established at the pseudogap energy (Fig. 4b).

To separate the nematic and smectic contributions in $Z(\mathbf{r}, e)$, we examine the spatial fluctuations of each ordering tendency by defining coarse-grained $\mathbf{r}$-space order parameter fields $O_n^Q(\mathbf{r}, e)$ and $O_s^Q(\mathbf{r}, e)$ using the coarsening length scales $1/\Lambda_n$ and $1/\Lambda_s$ shown as red and blue circles in Fig. 2 (Supplementary Information section VI). The resulting $O_n^Q(\mathbf{r}, e)$ and $O_s^Q(\mathbf{r}, e)$ (movies in the Supplementary Information) show that $O_s^Q(\mathbf{r}, e)$ spatially fluctuates wildly in the entire energy range whereas the spatial fluctuation of $O_n^Q(\mathbf{r}, e)$ rapidly subsides as it approaches $e \approx 1$. This dramatic difference is summarized in plots of the correlation lengths $\xi_n(e)$ and $\xi_s(e)$ extracted from $O_n^Q(\mathbf{r}, e)$ and $O_s^Q(\mathbf{r}, e)$ (Fig. 4b), wherein $\xi_n(e)$ diverges to the FOV size at $e \approx 1$. The representative spatial images of $O_n^Q(\mathbf{r}, e = 1)$ and $O_s^Q(\mathbf{r}, e = 1)$ in Fig. 4c and d show how distinct are the spatial structures of $O_n^Q(\mathbf{r}, e)$ and $O_s^Q(\mathbf{r}, e)$.

Our results also provide a new perspective which, by using $C_2$ symmetry as a common theme, may help to unify the understanding of angle-resolved photoemission (ARPES), neutron scattering and spectroscopic-imaging STM studies of broken electronic symmetries within the pseudogap phase. ARPES reveals spontaneous dichroism of the $\mathbf{k} = (1, 0)\,\pi/a_0$ and $\mathbf{k} = (0, 1)\,\pi/a_0$ states[24], exhibiting $C_2$ symmetry because the opposite sign of the effect occurs at $\mathbf{k} = (1, 0)\,\pi/a_0$ and $\mathbf{k} = (0, 1)\,\pi/a_0$ (ref. 24). These excitations, which are probably magnetic, appear at $T^*$ in $Bi_2Sr_2CaCu_2O_{8+\delta}$. The unusual magnetic order detected by polarized neutron diffraction at the Bragg peak[25,26] consists of magnetic moments of about $0.1\mu_B$ (where $\mu_B$ is the Bohr magneton) exhibiting $C_2$ symmetry. These intra-unit-cell signals appear at $T^*$ in both $YBa_2Cu_3O_{6+x}$ (ref. 25) and $HgBa_2CuO_{4+\delta}$ (ref. 26). Our studies reveal intra-unit-cell, $C_2$ symmetric excitations at the pseudogap energy and that these effects are associated primarily with electronic inequivalence at the two O sites within the $CuO_2$ unit cell. Given the many common characteristics observed by these diverse techniques, it is reasonable to consider whether ARPES, neutron diffraction and spectroscopic-imaging STM are detecting the same excitations with the same broken symmetries. If so, the pseudogap excitations of underdoped copper oxides would represent weakly magnetic states at the O sites within each $CuO_2$ unit cell, the electronic structure of which breaks $C_4$ symmetry. Then, the electronic symmetry breaking that occurs on entering the pseudogap phase would be due to the electronic nematic state visualized here, for the first time to our knowledge (Figs 3 and 4). Finally, the nematicity found in electronic transport[27], thermal transport[28] and the spin excitation spectrum[29] of $YBa_2Cu_3O_{6+x}$ could then occur because the Ising domains of $O_n^Q$ ( **r**, $e$ ) become aligned by the strong orthorhombicity of its crystal structure[30].

**Acknowledgements:** We are grateful to P. Abbamonte, D. Bonn, J.C. Campuzano, D.M. Eigler, E. Fradkin, T. Hanaguri, W. Hardy, S. Kivelson, A.P. Mackenzie, M. Norman, B. Ramshaw, S. Sachdev, G. Sawatzky, H. Takagi, J. Tranquada, and J. Zaanen, for helpful discussions and communications. Theoretical studies were supported by NSF DMR-0520404 to the Cornell Center for Materials Research. Experimental studies are supported by the Center for Emergent Superconductivity, an Energy Frontier Research Center, headquartered at Brookhaven National Laboratory and funded by the U.S. Department of Energy, under DE-2009-BNL-PM015. A.S. acknowledges support from the US Army Research Office. MJL, JCD and EAK thank KITP for its hospitality. JCD acknowledges gratefully the hospitality and support of the Physics and Astronomy Department at the University of British Columbia, Vancouver, BC, Canada.

**Figure 1 CuO$_2$ Electronic Structure and $\omega \approx \Delta_1$ Pseudogap States**

a. Schematic of the spatial arrangements of CuO$_2$ electronic structure with Cu sites and $d_{x^2-y^2}$ orbitals indicated in blue and O sites and 2$p_\sigma$ orbitals in yellow. $E_F$, Fermi energy. The inset shows the approximate energetics of the band structure when such a charge-transfer insulator is doped by removing electrons from the O atoms (the respective bands are indicated by the same colours as in the CuO$_2$ schematic). The 'real' part of any **q**-space electronic structure at the Bragg wavevector $\operatorname{Re}\tilde{M}(\mathbf{Q},\omega)$ is defined throughout this paper as being in phase with the Cu lattice (and therefore even about each Cu site). This definition is shown schematically as modulations in $\tilde{M}(\mathbf{r},\omega)$ which would contribute to $\operatorname{Re}\tilde{M}(\mathbf{Q},\omega)$ at the two Bragg wavevectors **Q**$_x$ and **Q**$_y$. DOS, density of states.

b. Schematic copper-oxide phase diagram. Here $T_c$ is the critical temperature circumscribing a 'dome' of superconductivity, $T_\phi$ is the maximum temperature at which phase fluctuations are detectable within the pseudogap phase, and $T^*$ is the approximate temperature at which the pseudogap phenomenology first appears.

c. The two distinct classes of excitations as identified by multiple spectroscopies in underdoped copper-oxides (reproduced with permission from ref. 3) as a function of hole-density $p$. The excitations to energies $\Delta_1(p)$ are referred to as the 'pseudogap states' both because $\Delta_1(p)$ tracks $T^*(p)$ and because they exist unchanged in both the pseudogap and superconducting phases. Those excitations to energies $\omega < \Delta_0(p)$ can be associated with the Bogoliubov quasiparticles[6–8].

d. Evolution of the spatially averaged tunnelling spectra of Bi$_2$Sr$_2$CaCu$_2$O$_{8+\delta}$ with diminishing $p$, here characterized by $T_c(p)$. The energies $\Delta_1(p)$ (blue dashed line) are easily detected as the pseudogap edge while the energies $\Delta_0(p)$ (red dashed line) are more subtle but can be identified by the correspondence of the "kink" energy[4] with the extinction energy of Bogoliubov quasiparticles, following the procedure in ref. 8.

**Figure 2. Imaging the Spatial Symmetries of the $\omega \approx \Delta_1$ Pseudogap States**

a. Spatial image ($R$-map[5]) of the $Bi_2Sr_2CaCu_2O_{8+\delta}$ pseudogap states $\omega \approx \Delta_1$ at $T \approx 4.3$ K for an underdoped sample with $T_c = 35$ K. The inset shows the Fourier transform upon which the inequivalent Bragg vectors $\mathbf{Q}_x = (1, 0) 2\pi / a_0$ and $\mathbf{Q}_y = (0, 1) 2\pi / a_0$ are identified by red arrows and circles. The inequivalent wavevectors $\mathbf{S}_x = (\sim 3/4, 0) 2\pi / a_0$ and $\mathbf{S}_y = (0, \sim 3/4) 2\pi / a_0$ are identified by blue arrows and circles.

b. Spatial image ($R$-map[5]) of the $Bi_2Sr_2CaCu_2O_{8+\delta}$ pseudogap states $\omega \approx \Delta_1$ at T $\approx 55$ K for the same sample with $T_c = 35$ K. Again, the inset shows the Fourier transform with the inequivalent Bragg vectors $\mathbf{Q}_x = (1, 0) 2\pi / a_0$ and $\mathbf{Q}_y = (0, 1) 2\pi / a_0$ identified by red arrows and $\mathbf{S}_x = (\sim 3/4, 0) 2\pi / a_0$ and $\mathbf{S}_y = (0, \sim 3/4) 2\pi / a_0$ identified by blue arrows and circles. The phenomenology of the $\omega \approx \Delta_1$ pseudogap states, especially their broken spatial symmetries, appear indistinguishable whether in the superconducting phase (a) or in the pseudogap phase (b).

c, A schematic representation of how electronic contributions from multiple sites within the $CuO_2$ unit cell could lead to global electronic nematicity in the copper oxides. Here the two O sites are labelled using different colours to represent the inequivalent electronic structure at those locations within each unit cell.

**Figure 3 Nematic Ordering and O-site specificity of $\omega \approx \Delta_1$ Pseudogap States**

a. Topographic image $T(\mathbf{r})$ of the $Bi_2Sr_2CaCu_2O_{8+\delta}$ surface. The inset shows that the real part of its Fourier transform $\mathrm{Re}\,T(\mathbf{q})$ does not break $C_4$ symmetry at its Bragg points because plots of $T(\mathbf{q})$ show its values to be indistinguishable at $\mathbf{Q}_x = (1, 0)\, 2\pi/a_0$ and $\mathbf{Q}_y = (0, 1)\, 2\pi/a_0$. Importantly, this means that neither the crystal nor the tip used to image it (and its $Z(\mathbf{r}, \omega)$ simultaneously) exhibits $C_2$ symmetry (Supplementary Information section V). The bulk incommensurate crystal supermodulation is seen clearly here; as always, it is at 45° to, and therefore is the mirror plane between, the *x* and *y* axes. For this symmetry reason it has no influence on the electronic nematicity discussed in this paper.

b. The $Z(\mathbf{r}, e = 1)$ image measured simultaneously with $T(\mathbf{r})$ in **a**. The inset shows that the Fourier transform $Z(\mathbf{q}, e = 1)$ does break $C_4$ symmetry at its Bragg points because $\mathrm{Re}\,\tilde{Z}(\mathbf{Q}_x, e) \neq \mathrm{Re}\,\tilde{Z}(\mathbf{Q}_y, e)$. This means that, on average throughout the FOV of **a** and **b**, the modulations of $Z(\mathbf{r}, \omega \approx \Delta_1)$ that are periodic with the lattice have different intensities along the *x* axis and along the *y* axis. This is a priori evidence for electronic nematicity in the pseudogap states $\omega \approx \Delta_1$.

c. The value of $O_n^Q(e)$ defined in equation (3) computed from $Z(\mathbf{r}, e)$ data measured in the same FOV as **a** and **b**. Its magnitude is low for all $\omega < \Delta_0$ and then rises rapidly to become well established near $e \approx 1$ or $\omega \approx \Delta_1$. Thus the quantitative measure of intra-unit-cell electronic nematicity established in equations (1) and (3) reveals that the pseudogap states in this FOV of a strongly underdoped $Bi_2Sr_2CaCu_2O_8$ sample are nematic.

d. Topographic image $T(\mathbf{r})$ from the region identified by a small black box in **a**. It is labelled with the locations of the Cu atom plus both the O atoms within each $CuO_2$ unit cell (labels shown in the inset). Overlaid is the location and orientation of a Cu and four surrounding O atoms using a representation similar to that of Fig. 2c.

e. The simultaneous $Z(\mathbf{r}, e = 1)$ image in the same FOV as **d** (the region identified by small white box in **b**) showing the same Cu and O site labels within each unit cell (see inset). Thus the physical locations at which the nematic measure $O_n^R(e)$ of equation (4) is evaluated are labelled by the dashes. Overlaid is the location and orientation of a Cu atom and four surrounding O atoms using a representation similar to that of Fig. 2c.

f. The value of $O_n^R(e)$ in equation (4) computed from $Z(\mathbf{r}, e)$ data measured in the same FOV as **a** and **b**. As in **c**, its magnitude is low for all $\omega < \Delta_0$ and then rises rapidly to become well established at $e \approx 1$ or $\omega \approx \Delta_1$. If the function in equation (4) is evaluated using the Cu sites only, the nematicity is about zero (black diamonds), as it must be. This independent quantitative measure of intra-unit-cell electronic nematicity $O_n^R(e)$ again shows that the pseudogap states are strongly nematic and, moreover, that the nematicity is due primarily to electronic inequivalence of the two O sites within each unit cell.

**Figure 4. Rapid Increase of Correlation Length of Nematicity at $\omega \approx \Delta_1$**

a. A large FOV $T(\mathbf{r})$ image which preserves $C_4$ symmetry.

b. Correlation lengths $\xi_n(e)$ for the nematic ordering $O_n^Q(\mathbf{r}, e)$ (red solid diamonds), and $\xi_s(e)$ for the possibly smectic ordering $O_s^Q(\mathbf{r}, e)$ (blue solid squares). (See Supplementary Information section VI for an evaluation of $O_n^Q(\mathbf{r}, e)$ and $O_s^Q(\mathbf{r}, e)$.) The coarsening length scales $1/\Lambda_n$ and $1/\Lambda_s$ are one-third that of typical spatial variations, as determined from the $3\sigma$ radius of the respective peaks. The correlation lengths are determined from the full-width at half-maximum of the spatial auto-correlation functions. Comparison of the nematic order parameter $O_n^Q(e)$ evaluated from equation (3) (open diamonds) and smectic order parameter $O_s^Q(e)$) from equation (6) (open squares) shows that $O_s^Q(e)$ has low magnitude and is energy independent, whereas $O_n^Q(e)$ rises rapidly to become well established at the pseudogap energy.

c. Image of $O_n^Q(\mathbf{r}, e = 1)$ (Supplementary Information section VI) from the same FOV showing that $\xi_n(e)$ has diverged to the size of the image. Thus, if there are Ising nematic domains (as there should be), they must be larger than 0.025 micrometres square. We note that the spatial resolution is limited by the cut-off scales shown in the figure.

d. Images of $O_s^Q(\mathbf{r}, e = 1)$ (Supplementary Information section VI) from the same FOV as a showing that $\xi_s(e)$ is spatially disordered with very short correlation length; this is equally true at all energies. The evolution of $O_n^Q(\mathbf{r}, e)$ and $O_s^Q(\mathbf{r}, e)$ as a function of e are available in Supplementary movies 1 and 2, respectively.

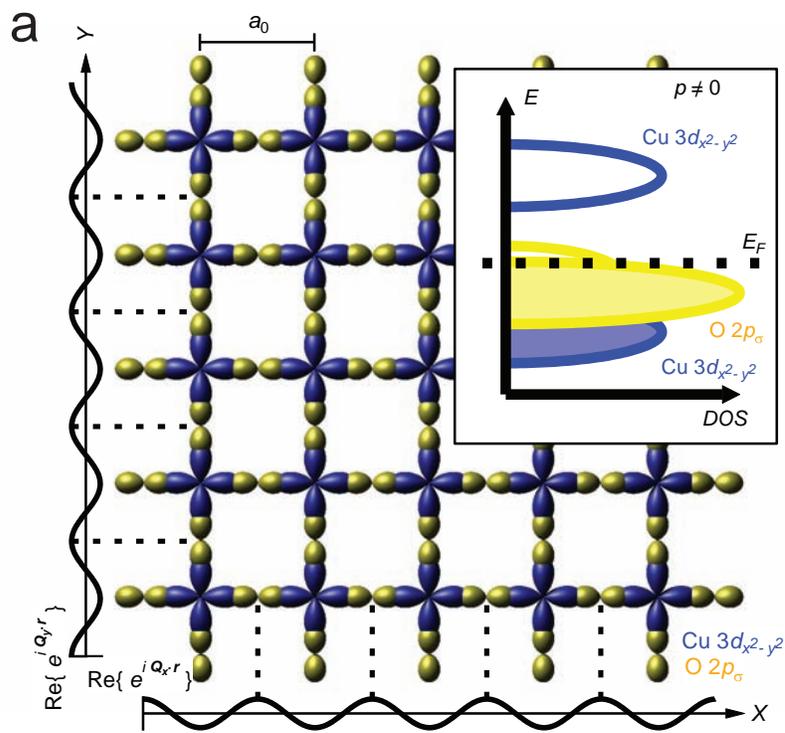
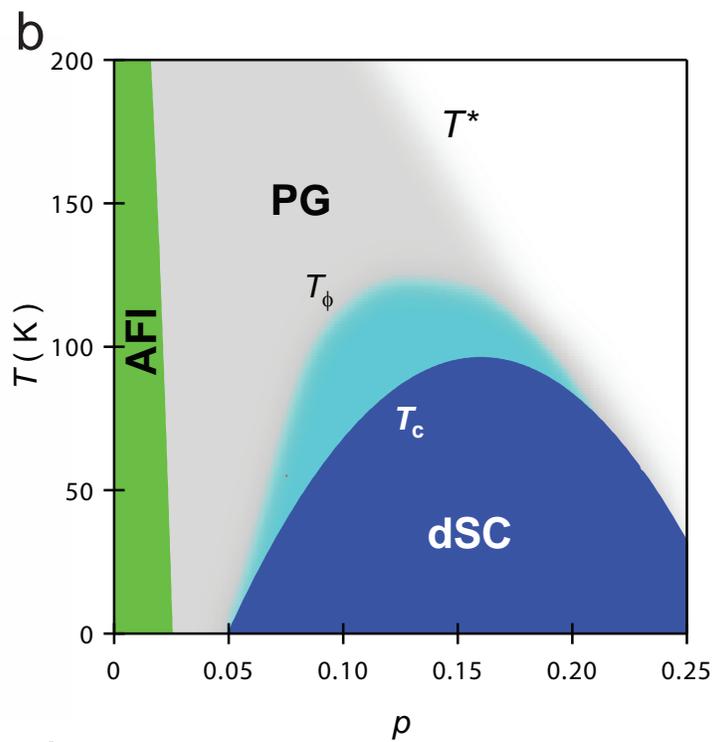
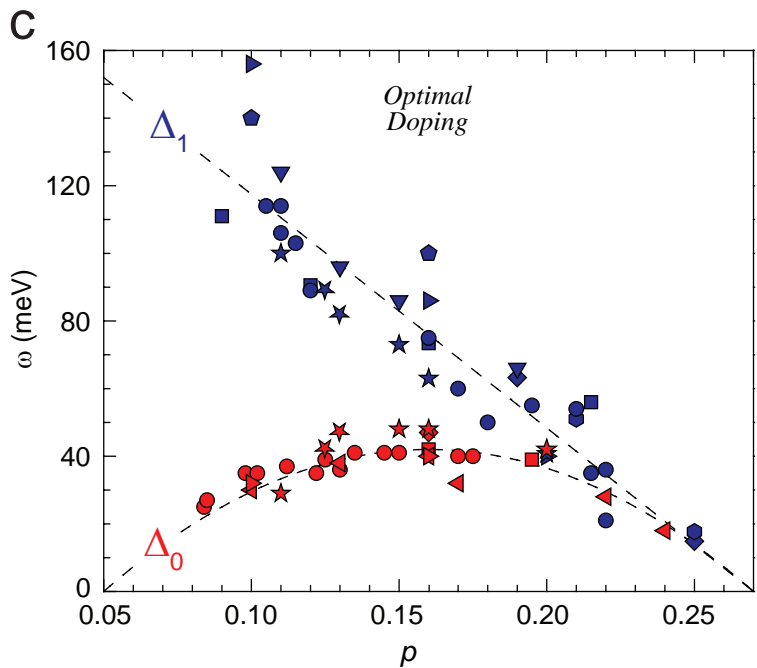
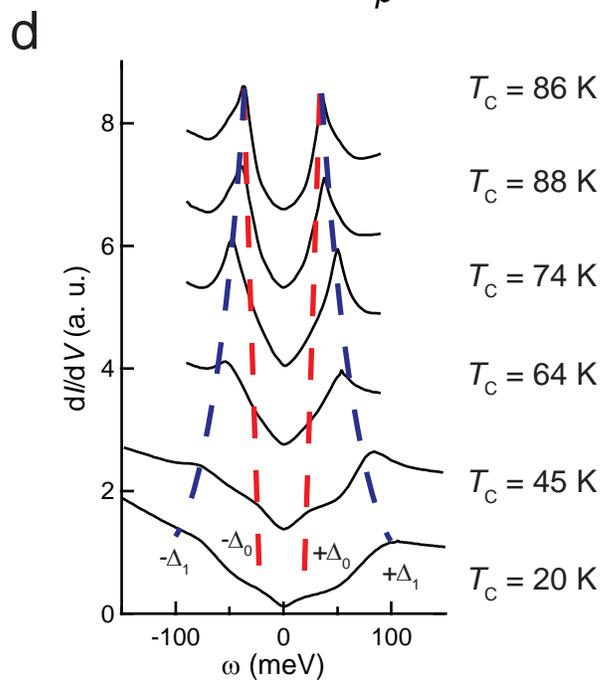

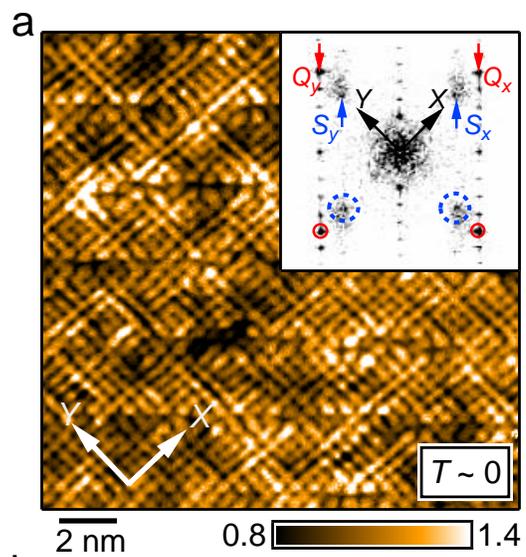
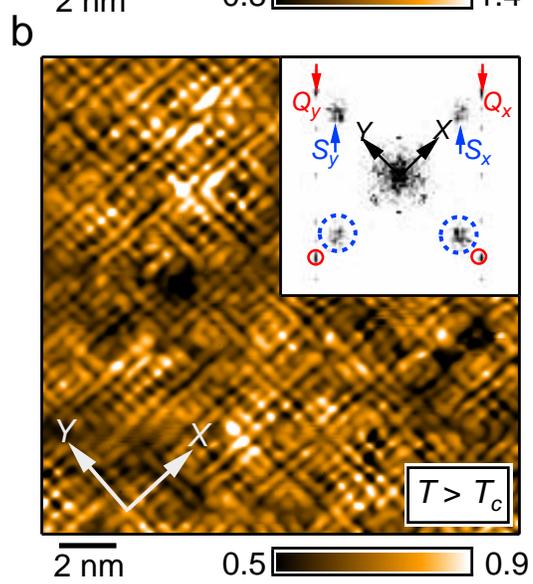
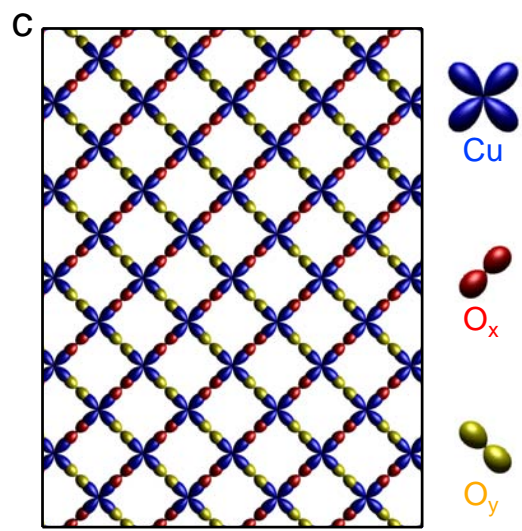

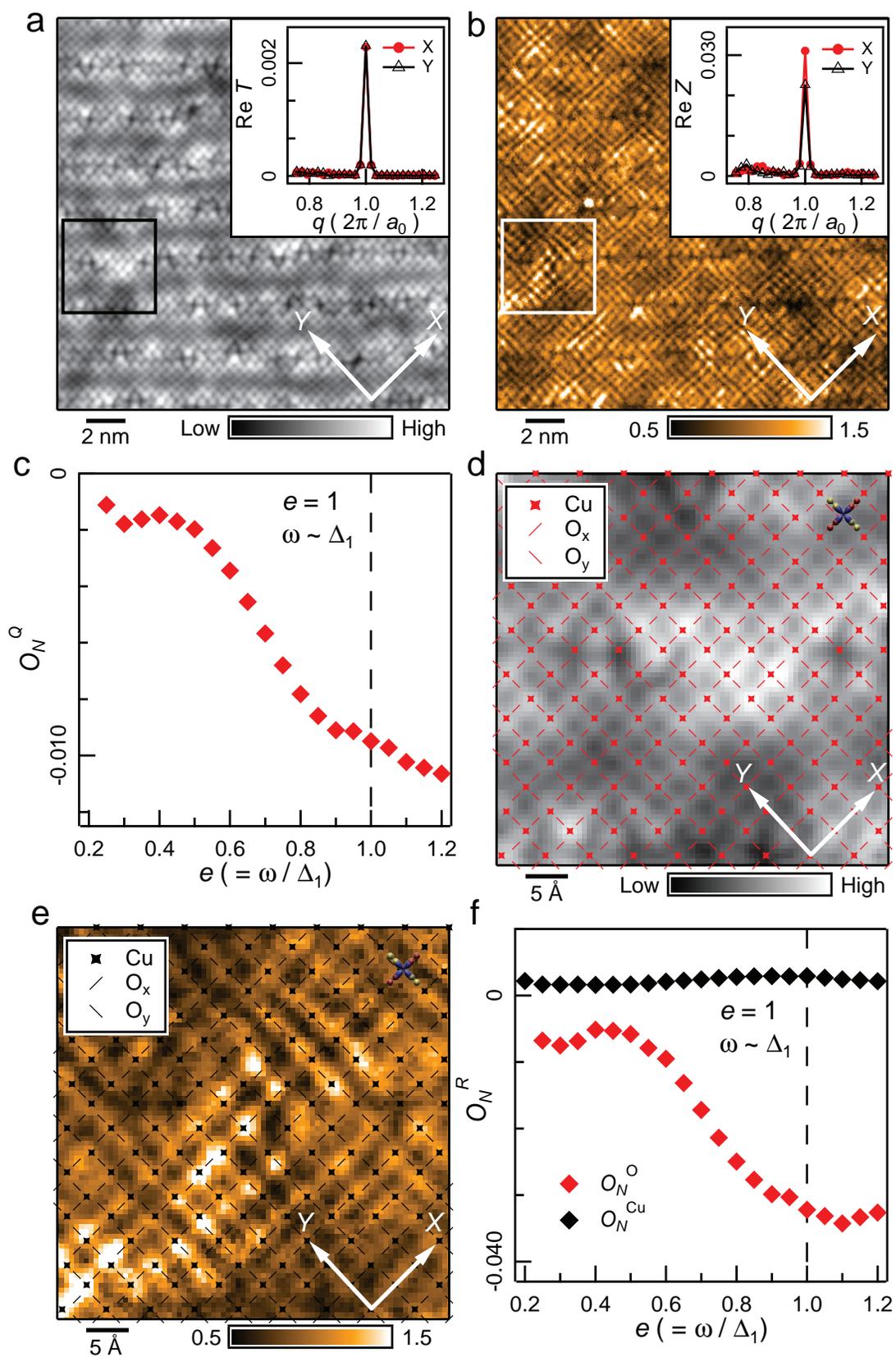

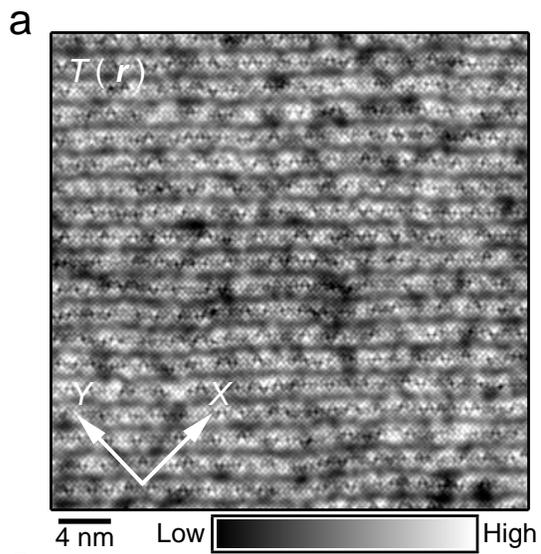
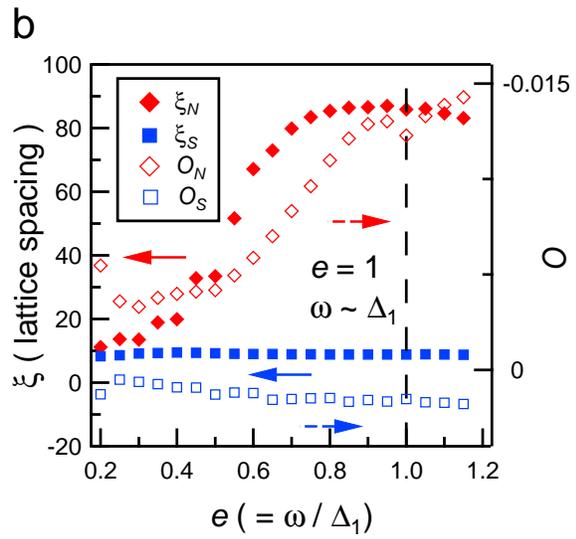
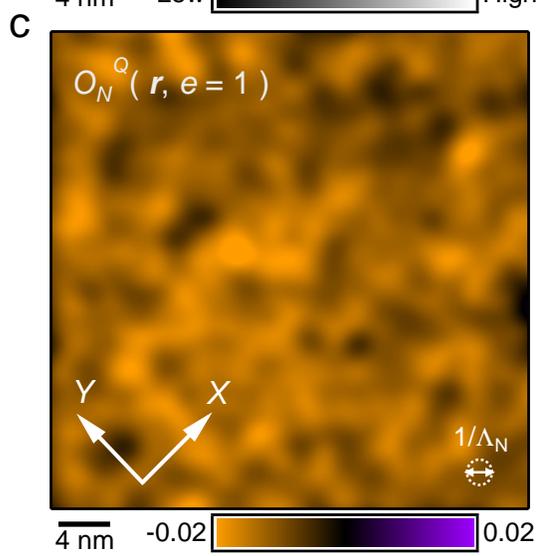
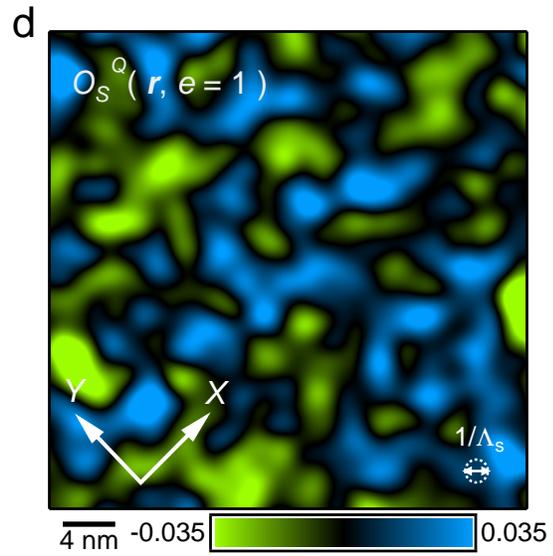

Supporting Online Material for

# Electronic Nematic Ordering of the Intra-unit-cell Pseudogap States in Underdoped Bi$_2$Sr$_2$CaCu$_2$O$_{8+\delta}$

M. J. Lawler, K. Fujita, Jhinhwan Lee, A. R. Schmidt, Y. Kohsaka, Chung Koo Kim, H. Eisaki, S. Uchida, J. C. Davis, J. P. Sethna, and Eun-Ah Kim

(I) **Use of** $Z(\mathbf{r},\omega)$ **and** $Z(\mathbf{r},e)$ **for Electronic Structure Studies**

Direct visualization of whether the copper-oxide pseudogap states exhibit intra-unit-cell electronic nematicity should be possible using subatomic resolution SI-STM. If used correctly, this technique can access simultaneously **r**-space and **k**-space electronic structure for both filled and empty states (Ref.'s 6-9, 28-30). However, care is required to avoid serious systematic errors which occur because the tunneling current is

$$I(\mathbf{r},z,V) = f(\mathbf{r},z)\int_0^{eV} N(\mathbf{r},\omega)d\omega \quad \text{(S1)}$$

where $z$ is the tip-surface distance, $V$ the tip-sample bias voltage, $N(\mathbf{r},\omega)$ the sample's local-density-of-states at location **r** and energy $\omega$, while $f(\mathbf{r},z)$ contains both effects of tip elevation and of spatially-dependent tunneling matrix elements (Ref.'s 6-9). The differential tunneling conductance $dI/dV(\mathbf{r},\omega=eV) \equiv g(\mathbf{r},\omega=eV)$ is then related to $N(\mathbf{r},\omega)$ by

$$g(\mathbf{r},\omega=eV) = \frac{eI_S}{\int_0^{eV_S} N(\mathbf{r},\omega')d\omega'} N(\mathbf{r},\omega) \quad \text{(S2)}$$

where $V_S$ is the junction-formation bias voltage. Thus when the electronic structure (and therefore $\int_0^{eV_S} N(\mathbf{r},\omega')d\omega'$) are heterogeneous as in underdoped copper-oxides (Ref.'s 6-9), $g(\mathbf{r},\omega=eV)$ cannot be used to measure $N(\mathbf{r},\omega)$. Nevertheless, by using the ratio

$$Z(\mathbf{r},\omega) \equiv \frac{g(\mathbf{r},\omega=+eV)}{g(\mathbf{r},\omega=-eV)} \equiv \frac{N(\mathbf{r},+\omega)}{N(\mathbf{r},-\omega)} \quad \text{(S3)}$$

these potentially severe systematic errors are cancelled manifestly (Ref.'s 6-9). In this case, $Z(\mathbf{r},\omega)$ is well defined as the ratio of the probability to inject an electron to the probability to extract one at a given ω. Another challenge in Bi$_2$Sr$_2$CaCu$_2$O$_{8+\delta}$ is the random nanoscale variation of $\Delta_1(\mathbf{r})$ which causes the pseudogap states to be detected at different bias voltages at different sample locations. However, this effect is mitigated by scaling the energy $\omega$ at each **r** by the pseudogap magnitude $\Delta_1(\mathbf{r})$ at the same location, thus defining a new reduced energy scale $e = \omega/\Delta_1(\mathbf{r})$ and

$$Z(\mathbf{r},e) \equiv Z(\mathbf{r},\omega/\Delta_1(\mathbf{r})) \quad \text{(S4)}$$

Then, in all such $Z(\mathbf{r},e)$ data, the $\omega \sim \Delta_1$ pseudogap states occur together at e=1 (Ref. 8).

## (II) Correcting Picometer-scale Distortions in Topography T(r) and Electronic Structure Z(r,ω)

Here we describe how to use the topograph to correct for the picometer scale drift of the tip location due to both piezoelectric mechanical creep and mK temperature variations over the approximately 1 week required for each Z(**r**,ω) data set to be acquired. A typical topograph shows variations at differing length scales associated with different physics: modulations with wave vectors $\mathbf{Q}_x$ and $\mathbf{Q}_y$ for Bi (and thus Cu) lattice and super-lattice modulation with wave vector $\mathbf{Q}_{sup}$, plus a slowly varying apparent "displacement" due to long-term and picometer scale piezoelectric drift. We define the slowly varying "displacement" field $\vec{u}(\mathbf{r})$ such that un-displaced positions $\mathbf{r} - \vec{u}(\mathbf{r})$ (which are the Lagrangian coordinates of elasticity theory) will form a perfect square lattice with Cu lattice locations $\vec{d}_{Cu} = 0$ within the unit cell. Now the topograph is expected to take the form

$$T(\mathbf{r}) = T_0 \left[ \cos(\mathbf{Q}_x \cdot (\mathbf{r} - \vec{u}(\mathbf{r}))) + \cos(\mathbf{Q}_y \cdot (\mathbf{r} - \vec{u}(\mathbf{r}))) \right] + T_{sup} \cos(\mathbf{Q}_{sup} \cdot (\mathbf{r} - \vec{u}(\mathbf{r}))) + \ldots \quad (S5)$$

where ... represent other contributions such as impurities etc. That $\vec{u}(\mathbf{r})$ is slowly varying compared to the scale of the super-lattice modulation and the lattice is evident from the Fourier transform of the topograph. In order to extract the slow varying $\vec{u}(\mathbf{r})$ it is useful to introduce a coarsening length scale $1/\Lambda_u$ over which $\vec{u}(\mathbf{r})$ is roughly constant such that $\Lambda_u \ll |\mathbf{Q}_{sup}|, |\vec{\mathbf{Q}}_{x,y}|$. The Fourier transform of the topograph shows that we can quite safely choose fairly small $\Lambda_u$ since the lattice peak is quite sharp (Fig. 3a). Now consider

$$T_x(\mathbf{r}) = \sum_{\mathbf{r}'} T(\mathbf{r}') e^{-i\mathbf{Q}_x \cdot \mathbf{r}'} \left( \frac{\Lambda_u^2}{2\pi} e^{-\Lambda_u^2 |\mathbf{r}-\mathbf{r}'|^2/2} \right) \quad (S6)$$

the weighted average of $T(\mathbf{r}')e^{-i\mathbf{Q}_x \cdot \mathbf{r}'}$ over the length scale $1/\Lambda_u$. Since $\Lambda_u \ll |\mathbf{Q}_{sup}|, |\vec{\mathbf{Q}}_{x,y}|$, their contributions average out, leaving

$$T_x(\mathbf{r}) \approx (T_0/2) e^{-i\mathbf{Q}_x \cdot \vec{u}(\mathbf{r})} \quad (S7)$$

We made use of the fact that $\vec{u}(\mathbf{r}') \approx \vec{u}(\mathbf{r})$ for small $|\mathbf{r} - \mathbf{r}'| < \Lambda_u$. Similarly we can define

$$T_y(\mathbf{r}) = \sum_{\mathbf{r}'} T(\mathbf{r}') e^{-i\mathbf{Q}_y \cdot \mathbf{r}'} \left( \frac{\Lambda_u^2}{2\pi} e^{-\Lambda_u^2 |\mathbf{r}-\mathbf{r}'|^2/2} \right) \approx (T_0/2) e^{-i\mathbf{Q}_y \cdot \vec{u}(\mathbf{r})} \quad (S8)$$

Hence we can extract the "displacement field" $\vec{u}(\mathbf{r})$ (much in the spirit of elasticity theory) and thus undo all effects of piezoelectric and/or thermal drift and also set the origin of the coordinate system such that $\vec{d}_{Cu} = 0$.

It turns out that the drift over the extent of a typical image is approximately one or two lattice-spacings out of >100 or less than 10 picometer per unit cell. This manifests itself through the phase $\mathbf{Q}_x \cdot \vec{u}(\mathbf{r})$ and $\mathbf{Q}_y \cdot \vec{u}(\mathbf{r})$, defined through an arctan2(x, y) function (with z = x + i y), which jumps by $2\pi$ in some regions. These phase jumps need to be removed to make $\vec{u}(\mathbf{r})$ a single valued quantity. Taking a derivative of the image to locate the jumps and adding $2\pi$ to $\vec{u}(\mathbf{r})$ where appropriate can perform this function. Finally, and importantly, the same geometrical transformation in Eqn.'s S7 and S8 are carried out on each Z(**r**,e) acquired simultaneously with the T(**r**) so that both are registered to each other while being rendered $a_o$ periodic. Without these procedures, the $O_N^Q(e)$ function of Eqn. 3 cannot be evaluated meaningfully.

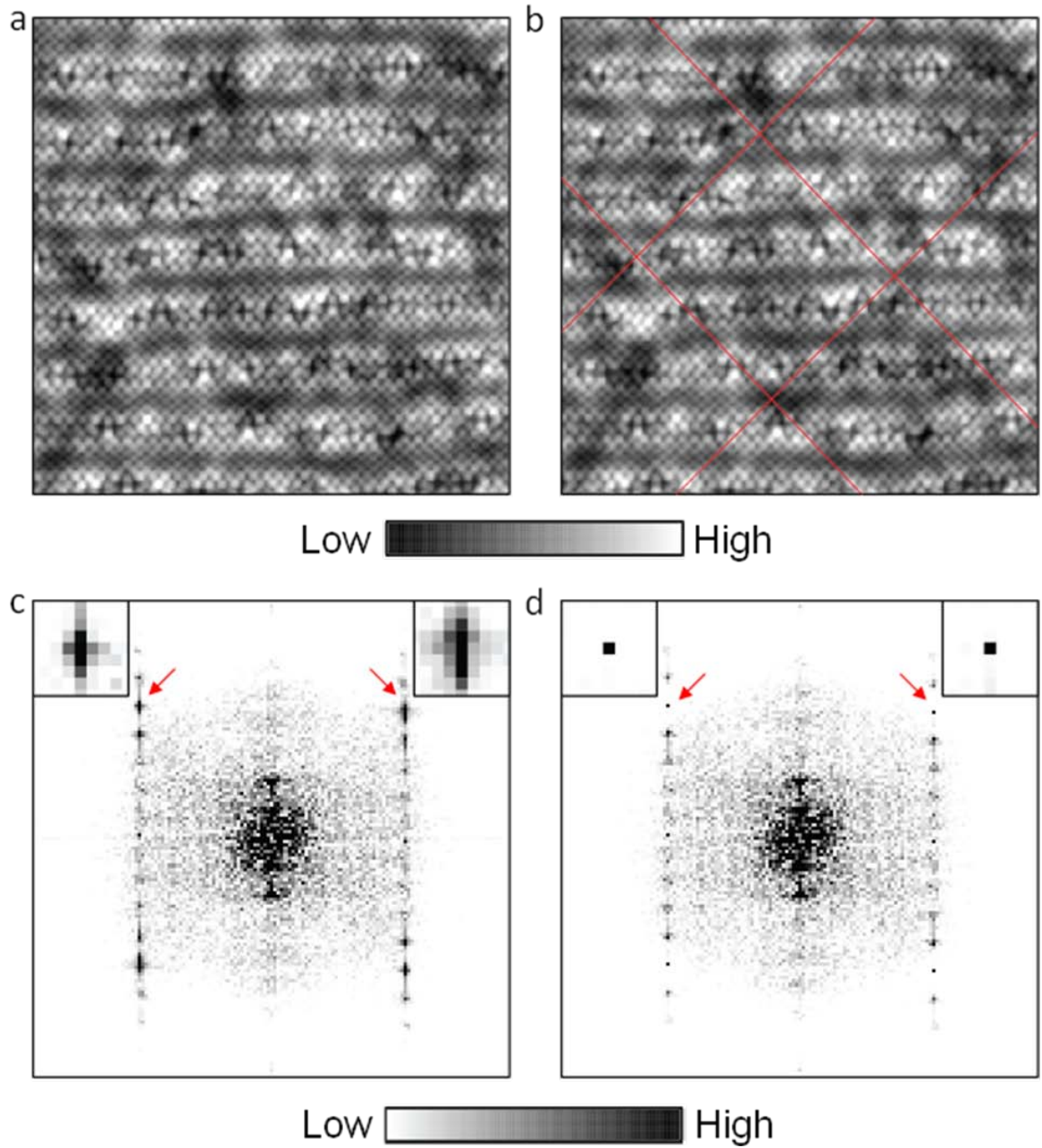

**Figure S1.** Topographic images of surface studied both before (a) and after (b) the distortion correction procedure. S1c shows the Fourier transform of the raw topograph while S1d shows the result of the correction; the Bragg peak (red arrows) has been reduced from ~10 pixels to a single pixel. Thus the topograph in b is $a_0$ periodic over the whole image, as required for meaningful use of Eqn. 3.

(III) **Repeatability of the observation of Intra-unit-cell Nematicity**

The same sequence of measurements as described in Fig. 3a-f were repeated on three different strongly underdoped $Bi_2Sr_2CaCu_2O_{8+\delta}$ samples ($T_c$ = 20K, 40K and 55K) using two different STM systems. The results are shown in Fig.'s S2a-f, S3a-f and S4a-f:

a. Topographic image T(**r**). Inset: The real part of its Fourier transform ReT(**q**) does not break $C_4$ symmetry at its Bragg points because plots of T(**q**) show it's values to be indistinguishable at $\mathbf{Q}_x=(1,0)2\pi/a_0$ and $\mathbf{Q}_y=(0,1)2\pi/a_0$. Neither the crystal nor the tip used to image it (and its Z(**r**,ω) simultaneously) exhibits $C_2$ symmetry.
b. The Z(**r**,e=1) image measured simultaneously with T(**r**) in a. The inset shows that the Fourier transform Z(**q**,e=1) does break $C_4$ symmetry at its Bragg points because $\operatorname{Re}\tilde{Z}(\mathbf{Q_x},e=1) \neq \operatorname{Re}\tilde{Z}(\mathbf{Q_y},e=1)$. On average throughout the FOV of a,b, the modulations of Z(**r**,ω~$\Delta_1$) that are lattice-periodic have different intensities along the x-axis and along the y-axis i.e. electronic nematicity exists in the pseudogap states ω~$\Delta_1$.
c. The value of $O_N^Q(e)$ from Eqn. 3 computed from Z(**r**,e) data measured simultaneously with a and b. Its value is low for ω<$\Delta_0$ and then rises rapidly to become well established at e~1 or ω~$\Delta_1$. Thus the quantitative measure of intra-unit-cell electronic nematicity established in Eqn. 1 reveals that the pseudogap states in this FOV are clearly nematic.
d. Topographic image T(**r**) from the region identified by a small box in a. It is labeled with the locations of the Cu atom plus both the O atoms within each $CuO_2$ unit cell (labels shown in the inset). Overlaid is the location and orientation of a Cu and four surrounding O atoms using a similar representation as in c.
e. The simultaneous Z(**r**,e=1) image in same FOV as d showing same Cu and O site labels within each unit cell (see inset). Thus the physical locations at which the nematic measure $O_N^R(e)$ of Eqn. 4 is always evaluated are labeled by the dashes.
f. The value of $O_N^R(e)$ in Eqn. 4 computed from Z(**r**,e) data measured in the same FOV as a and b. As in c, its magnitude is low for all ω<$\Delta_0$ and then rises rapidly to become well established at e~1 or ω~$\Delta_1$. If the function in Eqn. 4 is evaluated using the Cu sites only the nematicity is ~zero (black diamonds). Moreover $O_N^Q(e)$ and $O_N^R(e)$ should not be equal to each other since the former also involves effects at the Cu atoms. Thus $O_N^R(e)$ reveals that the pseudogap states are strongly nematic and, moreover, that the nematicity is due to electronic inequivalence of the two O sites within each unit cell.

In summary, the basic phenomenology of electronic nematicity is detected consistently by both $O_N^Q(e)$ and by $O_N^R(e)$ in 100% of the strongly underdoped samples studied. The only difference is that for some samples the nematicity is along the x-axis (Fig. S2) while for others it is along the y-axis (Fig.3 and S3,4); this is as would be expected for an Ising type nematic order parameter.

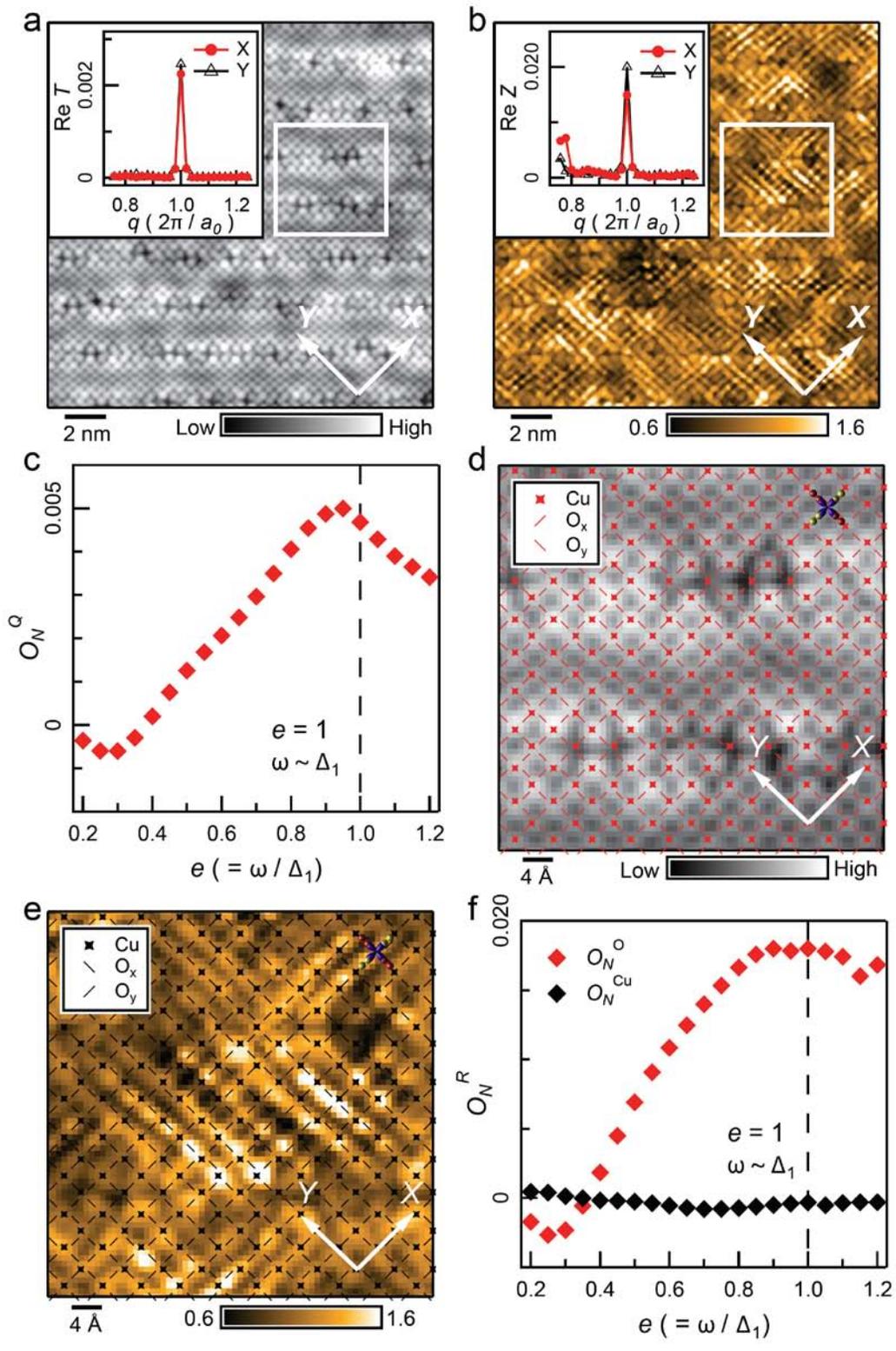

**Figure S2**

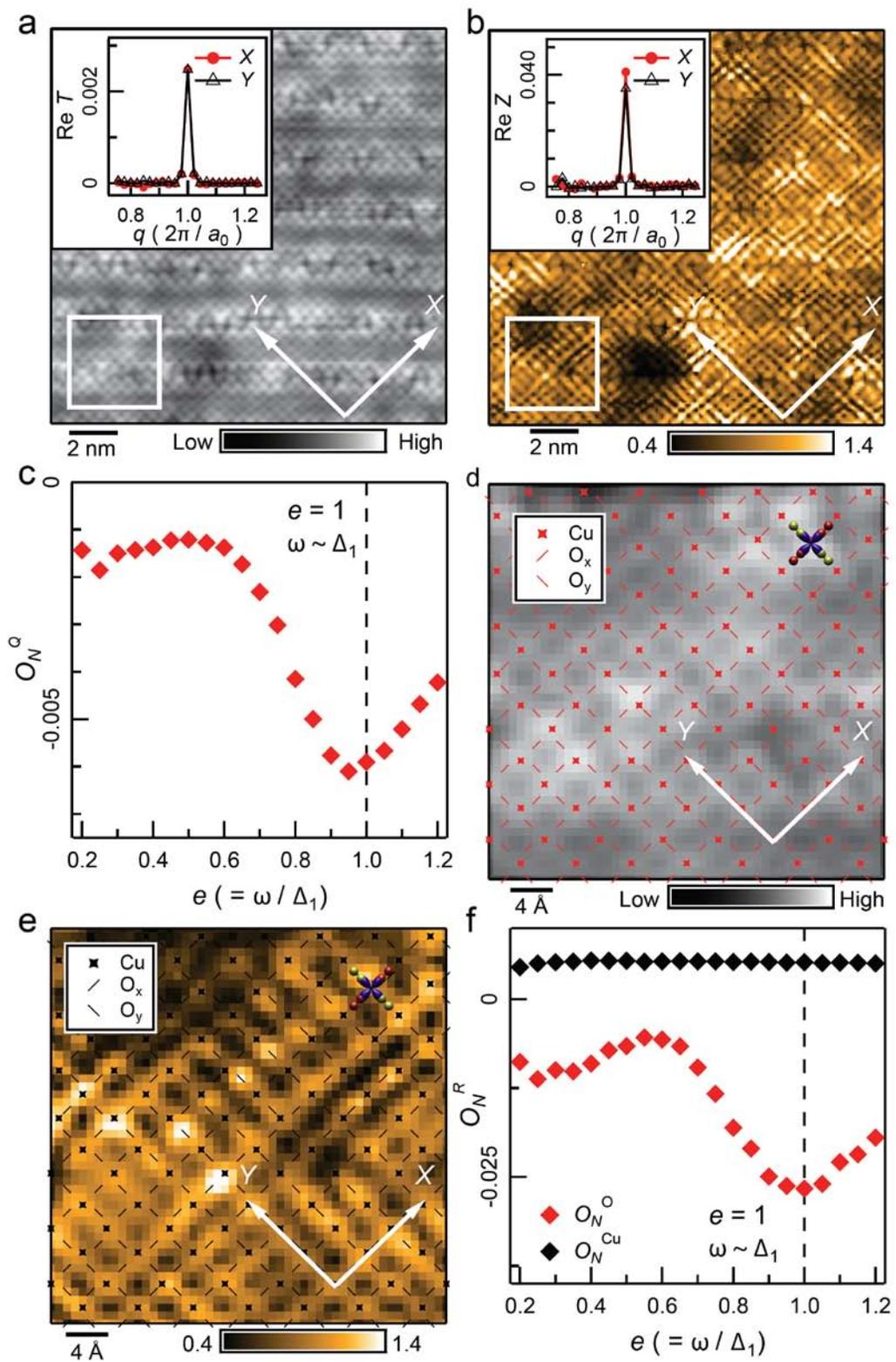

**Figure S3**

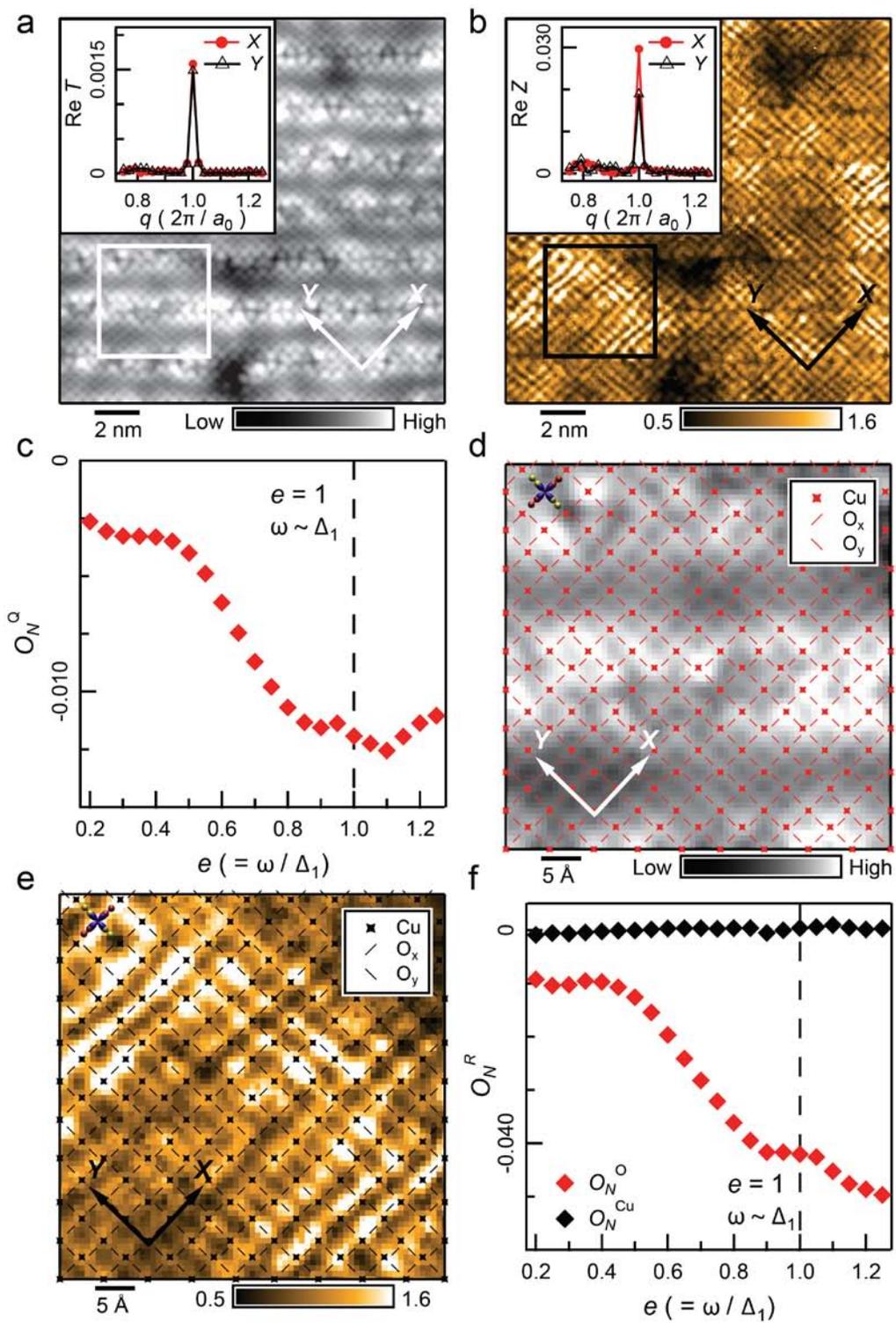

**Figure S4**

## (IV) Evolution of the unit-cell-average electronic structure with e.

Since the location of every CuO$_2$ unit cell is known with high prevision in the Z(**r**,e) data (here from Fig. 3) , one can determine the average electronic structure within a CuO$_2$ unit cell as a function of e. This result is shown in Fig. S5 where the average unit cell electronic structure is tiled in a 4X4 matrix to help in visualization of electronic structure evolution. We see clearly and directly that, within the unit cell, the electronic structure has C$_4$ symmetry at e=0.2 and evolves to strong C$_2$ symmetry at e=1.0.

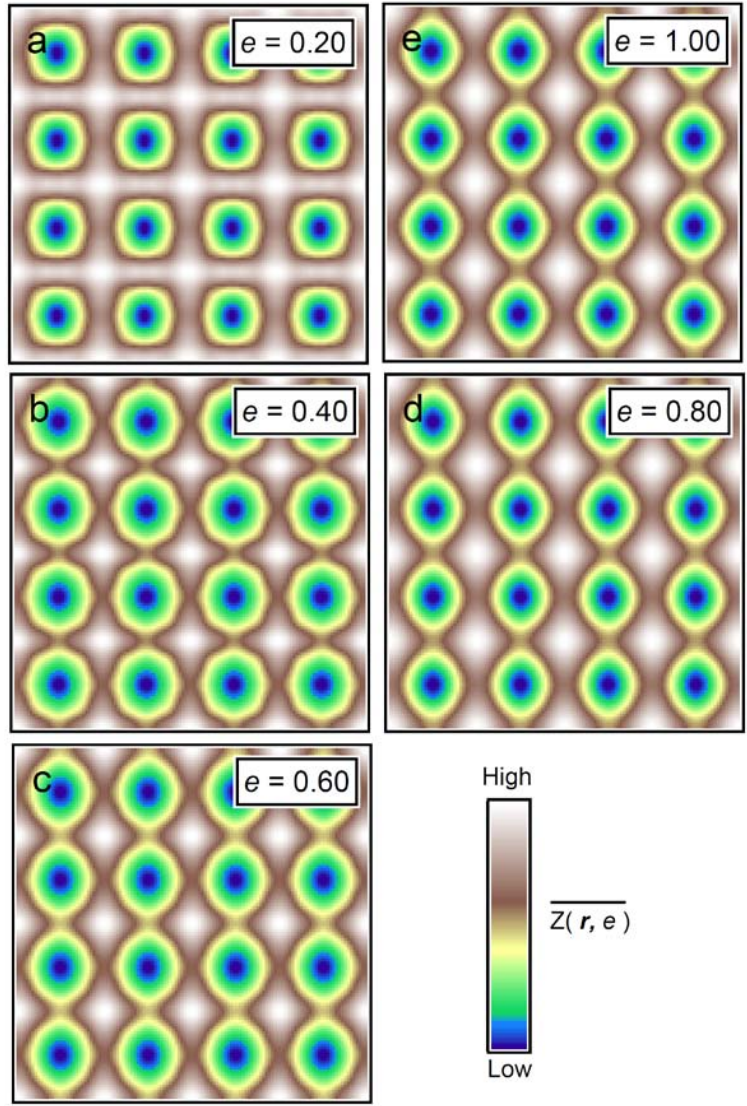

**Figure S5** The electronic structure of an average unit cell is measured as a function of e and then represented by tiling 4X4 copies of the average unit cell. We see directly the evolution away from the C$_4$ electronic structure at low e to strong C$_2$ electronic structure at e~1, within each unit cell.

**(V)  Exclusion of potential systematic errors**
A number of potential systematic errors have been excluded during these studies.

(i) One possible systematic error could be that the STM-tip shape breaks $C_4$ symmetry. However, we show that this is not the case in every data set described, for the following reasons. First, the topograph T(**r**) does not break $C_4$ symmetry in every case. If the tip geometry breaks any rotational symmetry, this is observed as a breaking of that symmetry in the geometry of the surface imaged by topography; however such signatures are demonstrably absent in every data set reported herein (Fig.'s 3a, S2a,S3a,S4a.) Second, the $C_2$ symmetry observed in Z(**r**,ω) is a strong function of the bias energy eV= ω and indeed is focused on the pseudogap energy (Fig. 3, S2,S3,S4,S5). This could not be the case if the phenomena reported were merely a geometrical characteristic of the tip/sample geometry which must be the same for every energy. Third, it is inconceivable that the geometry of four different tips in these independent experiments all happened to break the identical rotational symmetry and furthermore all happened to become aligned only with the x-axis or y-axis of each crystal.

(ii) Another possibility is that the procedures of raster scanning the tip over the surface causes the simultaneous T(**r**) and Z(**r**,ω) data sets to appear to break $C_4$ because of a $C_2$-symmetric process aligned with the raster direction. However, scan direction tests described in Fig. S6 demonstrate directly that this error does not exist because the actual data is demonstrably independent of the scan direction (except for the fact that there were necessarily fewer pixels when scanning along the a-axis Fig. S5c).

(iii) Another possibility is that the mechanical geometry of the STM scan head itself breaks $C_4$ symmetry. However since we used two completely different STM systems to carry out the studies: the data in Fig. 3, Fig S2 is from a physically distinct STM system than that in Fig. S3, S4. Thus, while the $C_2$ symmetry breaking effects are indistinguishable (Fig.'s 3, S2, S3, S4) the mechanical geometry of the STM scan heads are different, thus ruling out this effect.

(iv) A final possibility is that, after the drift corrections, the register of T(**r**) to Z(**r**,e) has a displacement in some specific direction which would result in a $C_4$ symmetry breaking due to the analysis procedure using Eqn. 3. However, we confirm directly that the cross correlation between every T(**r**) to Z(**r**,e) pair reported here is always peaked at the (0,0) point meaning that they are correctly registered to each other in that the origin is always at a Cu site in both data sets.

Thus we conclude that the potential systematic errors which could cause a spurious appearance of $C_2$ symmetry are positively ruled out in every case.

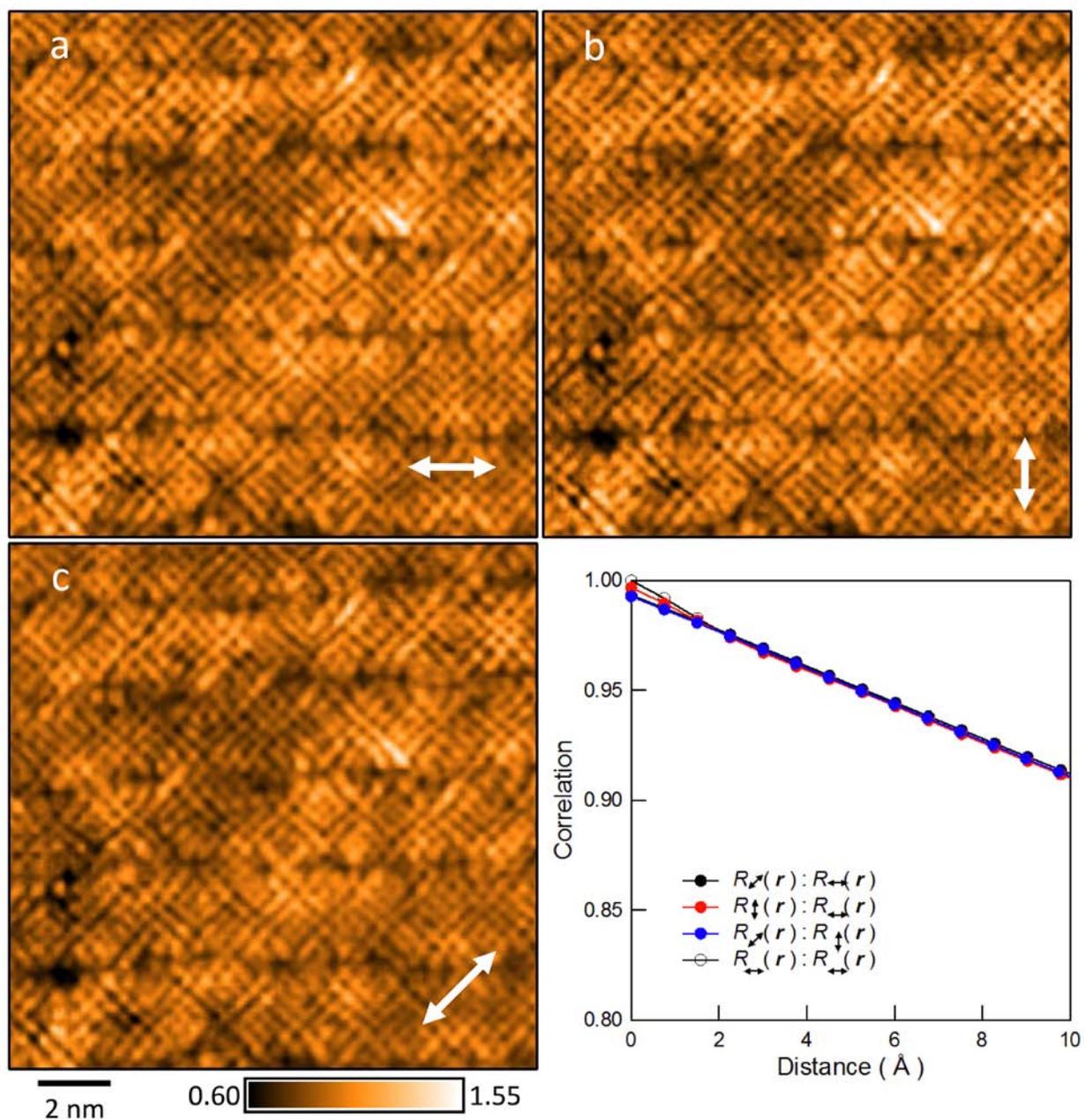

**Figure S6.** The electronic structure images of the pseudogap states near $\omega \sim \Delta_1$ is demonstrably identical for three different scan directions: a. parallel to the supermodulation, b. perpendicular to the supermodulation and c. parallel to the x-axis. The cross correlations of these images show that that they are virtually identical. Thus there is no scan-direction effect.

### (VI) Procedure for generating $O_N^Q(\mathbf{r},e)$ and $O_S^Q(\mathbf{r},e)$

To visualize the spatial structures contributing to $O_N^Q(e)$ we define a coarse grained field with a coarsening length scale $1/\Lambda_N$ which acts as an effective 'ultra violet' cutoff at the $\Lambda_N$ length scale (see circles around the Bragg peaks in Fig. 2). Then

$$\tilde{Z}(\mathbf{Q};\mathbf{r})_\Lambda \equiv \sum_{\mathbf{r}'} Z(\mathbf{r}')e^{i\mathbf{Q}\cdot\mathbf{r}'} f_\Lambda(\mathbf{r}'-\mathbf{r})$$

$$\approx \frac{1}{\sqrt{N}} \sum_{\mathbf{k}} \tilde{Z}(\mathbf{Q}-\mathbf{k})e^{i\mathbf{k}\cdot\mathbf{r}} e^{-\mathbf{k}^2/2\Lambda^2} \qquad (S9)$$

where $f_\Lambda(\mathbf{r}) \equiv (\Lambda^2/2\pi)e^{-\Lambda^2|\mathbf{r}|^2/2}$ is used to implement the cutoff at length scale $1/\Lambda$. For electronic nematicity, we set this cutoff to the 3σ radius of the Bragg peaks $\Lambda_N$ (Fig. 2). A local image of the nematicity $O_N^Q(\mathbf{r},e)$ is then given by:

$$O_N^Q(\mathbf{r},e) \propto \tilde{Z}(\mathbf{Q}_y;\mathbf{r})_{\Lambda_N} - \tilde{Z}(\mathbf{Q}_x;\mathbf{r})_{\Lambda_N} + \tilde{Z}(-\mathbf{Q}_y;\mathbf{r})_{\Lambda_N} - \tilde{Z}(-\mathbf{Q}_x;\mathbf{r})_{\Lambda_N} \Big|_e \qquad (S10)$$

with the normalization requiring the field of view (FOV) average to equal $O_N^Q$, i.e., $\langle O_N^Q(\mathbf{r}) \rangle = O_N^Q$. Similarly, a local image of any electronic smecticity $O_S^Q(\mathbf{r},e)$ is given by

$$O_S^Q(\mathbf{r},e) \propto \tilde{Z}(\mathbf{S}_y;\mathbf{r})_{\Lambda_S} - \tilde{Z}(\mathbf{S}_x;\mathbf{r})_{\Lambda_S} + \tilde{Z}(-\mathbf{S}_y;\mathbf{r})_{\Lambda_S} - \tilde{Z}(-\mathbf{S}_x;\mathbf{r})_{\Lambda_S} \Big|_e \qquad (S11)$$

Where, for the smectic order, we set this cutoff to the 3σ radius of the $\mathbf{S}_x$, $\mathbf{S}_y$ peaks $\Lambda_S$ (Fig. 2), and again normalized to $\langle O_S^Q(\mathbf{r}) \rangle = O_S^Q$. Simultaneous movies of $O_N^Q(\mathbf{r},e)$ and $O_S^Q(\mathbf{r},e)$ are presented.

### (VII) List of symbols

| | |
|---|---|
| $\Delta_1$ | Pseudogap maximum energy defined as in the Ref. 7 |
| $\Delta_0$ | Maximum energy of Bogoliubov QPI signature of Cooper pairing as defined as in the Ref. 7 |
| $T^*$ | Temperature below which pseudogap phenomena are observed |
| $\mathbf{Q}_x = (1, 0)2\pi/a_0$, $\mathbf{Q}_y = (0, 1)2\pi/a_0$ | Reciprocal lattice vectors for CuO$_2$ plane of Cu-Cu spacing $a_0$ |
| $\mathbf{S}_x \sim (3/4, 0)2\pi/a_0$, $\mathbf{S}_y \sim (0, 3/4)2\pi/a_0$ | Reciprocal lattice vectors for local smectic ordering |
| $M(\mathbf{r})$ | Real space image of arbitrary real-valued scalar quantity $M$ |
| $O_N[M] = \operatorname{Re}\tilde{M}(\mathbf{Q}_y) - \operatorname{Re}\tilde{M}(\mathbf{Q}_x)$ | Quantitative measure of amount of nematic ordering in $M(\mathbf{r})$ with respect to CuO$_2$ lattice |
| $\tilde{M}(\mathbf{q})$ | Complex-valued two-dimensional Fourier transform of $M(\mathbf{r})$ |
| $\overline{M}_{Cu}, \overline{M}_{O_x}, \overline{M}_{O_y}$ | Average of $M(\mathbf{r})$ across all Cu, O$_x$, and O$_y$ atoms, respectively. |
| $dI/dV(\mathbf{r},\omega=eV) = g(\mathbf{r}, \omega)$ | Spectroscopic imaging-scanning tunneling microscope derived conductance maps |
| $N(\mathbf{r},\omega)$ | Local density of states |
| $Z(\mathbf{r}, \omega = eV) = \dfrac{g(\mathbf{r},\omega)}{g(\mathbf{r},-\omega)} = \dfrac{N(\mathbf{r},\omega)}{N(\mathbf{r},-\omega)}$ | Conductance map ratio at opposite bias voltage polarities, which eliminates the systematic error due to STM constant current setup. |
| $e = \omega/\Delta_1(\mathbf{r})$ | Energy scaled according to local pseudogap maximum energy |
| $\tilde{Z}(\mathbf{q},e)$ | Two-dimensional Fourier transform of $Z(\mathbf{r},e)$, with rescaled energy. |
| $O_N^Q(e) = \dfrac{\operatorname{Re}\tilde{Z}(\mathbf{Q}_y,e) - \operatorname{Re}\tilde{Z}(\mathbf{Q}_x,e)}{\overline{Z}(e)}$ | Quantitative measure, in reciprocal space, of nematic ordering with respect to $CuO_2$ lattice vectors in $Z(\mathbf{r},e)$ |
| $O_N^R(e) = \sum_{\mathbf{R}} \dfrac{Z_x(\mathbf{R},e) - Z_y(\mathbf{R},e)}{\overline{Z}(e)N}$ | Quantitative measure, in real space, of nematic ordering with respect to $CuO_2$ lattice vectors in $Z(\mathbf{r},e)$ due to electronic inequivalence of intra unit-cell $O_x$ and $O_y$ sites. |
| $N$ | Number of unit cells |
| $\mathbf{R}$ | $CuO_2$ plane Bravais lattice vectors |
| $O_S^Q(e) = \dfrac{\operatorname{Re}\tilde{Z}(\mathbf{S}_y,e) - \operatorname{Re}\tilde{Z}(\mathbf{S}_x,e)}{\overline{Z}(e)}$ | Quantitative measure, in reciprocal space, of smectic ordering in $Z(\mathbf{r},eV)$. |
| $O_N^Q(\mathbf{r},e)$ | Real-space nematic order parameter field which averages to $O_N^Q(e)$. |
| $O_S^Q(\mathbf{r},e)$ | Real-space smectic order parameter field which averages to $O_S^Q(e)$. |
| $1/\Lambda_N$ | Coarsening length scale against which to measure spatial fluctuations in $O_N^Q(\mathbf{r},e)$. |
| $1/\Lambda_S$ | Coarsening length scale against which to measure spatial fluctuations in $O_S^Q(\mathbf{r},e)$. |
| $\xi_N(e)$ | Correlation length for spatial fluctuations of $O_N^Q(\mathbf{r},e)$ |
| $\xi_S(e)$ | Correlation length for spatial fluctuations of $O_S^Q(\mathbf{r},e)$ |